\documentclass[preprint]{aastex}

\usepackage{graphicx}
\usepackage{txfonts}
\usepackage{longtable}
\usepackage{lscape}
\usepackage{natbib}
\usepackage[usenames,dvipsnames,svgnames,table]{xcolor}
\shorttitle{The polarization of the Ca\,{\sc ii} IR triplet}
\shortauthors{\v{S}t\v{e}p\'an \& Trujillo Bueno}

\begin{document}

\title{The Hanle and Zeeman polarization signals of the solar Ca\,{\sc ii}~8542\,\AA\ line}

\author{Ji\v{r}\'i \v{S}t\v{e}p\'an\altaffilmark{1} \& Javier Trujillo Bueno\altaffilmark{2,3,4} }
\altaffiltext{1}{Astronomical Institute ASCR, Fri\v{c}ova 298, 251\,65 Ond\v{r}ejov, Czech Republic}
\altaffiltext{2}{Instituto de Astrof\'isica de Canarias, E-38205 La Laguna, 
Tenerife, Spain}
\altaffiltext{3}{Departamento de Astrof\'isica,
Universidad de La Laguna, E-38206 La Laguna, Tenerife, Spain}
\altaffiltext{4}{Consejo Superior de Investigaciones Cient\'ificas, Spain}

\begin{abstract}
We highlight the main results of a three-dimensional (3D) multilevel radiative transfer investigation about the solar disk-center polarization of the Ca {\sc ii} 8542 \AA\ line. First, we investigate the linear polarization due to the atomic level polarization produced by the absorption and scattering of anisotropic radiation in a 3D model of the solar atmosphere, taking into account the symmetry breaking effects caused by its thermal, dynamic and magnetic structure. Second, we study the contribution of the Zeeman effect to the linear and circular polarization. Finally, we show examples of the Stokes profiles produced by the joint action of atomic level polarization and the Hanle and Zeeman effects. We find that the Zeeman effect tends to dominate the linear polarization signals only in the localized patches of opposite magnetic polarity where the magnetic field is relatively strong and slightly inclined, while outside such very localized patches the linear polarization is often dominated by the contribution of atomic level polarization. We demonstrate that a correct modeling of this last contribution requires taking into account the symmetry breaking effects caused by the thermal, dynamic and magnetic structure of the solar atmosphere, and that in the 3D model used the Hanle effect in forward-scattering geometry (disk center observation) mainly reduces the polarization corresponding to the zero-field case. We emphasize that, in general, a reliable modeling of the linear polarization in the Ca {\sc ii} 8542 \AA\ line requires taking into account the joint action of atomic level polarization and the Hanle and Zeeman effects.         
\end{abstract}

\keywords{line: profiles --- polarization --- scattering 
--- radiative transfer --- Sun: chromosphere --- Stars: atmospheres}

\section{Introduction}

Information on the magnetic field of the solar chromosphere is encoded in the polarization of chromospheric spectral lines \citep[e.g.,][]{Harvey06}. A set of spectral lines whose polarization is considered to be of great diagnostic interest for investigating the magnetic activity of the solar chromosphere is the infrared (IR) triplet of Ca\, {\sc ii} at 8498\,\AA, 8542\,\AA\ and 8662\,\AA, with particular interest on its strongest line at 8542\,\AA\ whose line-core radiation encodes information on chromospheric layers around and slightly above the corrugated surface where the ratio $\beta$ of gas to magnetic pressure is unity \citep[e.g.,][]{Socas-Navarro+00,MansoTrujillo10,Cruz-Rodriguez+12}. In strongly magnetized regions (e.g., sunspots) the polarization of these lines is dominated by the Zeeman effect, and the ensuing Stokes profiles have been exploited to probe the magnetism of the observed regions through the application of non-LTE Stokes inversion techniques \citep[e.g.,][]{Socas-Navarro05}. Outside active regions, the circular polarization is still dominated by the Zeeman effect, but the linear polarization has, in general, significant contributions from scattering processes and the Hanle effect \citep[][]{MansoTrujillo10}.  

The spectral line polarization produced by the Zeeman effect results from the wavelength shifts between the $\pi$ ($\Delta{M}=0$) and $\sigma$ ($\Delta{M}={\pm}1$) transitions, with $M$ the magnetic quantum number of the atomic levels. In contrast, the physical origin of the scattering line polarization is the atomic polarization (population imbalance and quantum coherence among the Zeeman sublevels) of the upper and/or lower level of the line transition under consideration. Such atomic level polarization results from the absorption and scattering of anisotropic radiation, and the degree of anisotropy of the incident radiation field depends on the thermal and dynamic structure of the stellar atmosphere under study \citep[e.g.,][]{LL04}. In general, the polarization of solar spectral lines results from the joint action of atomic level polarization and the Hanle and Zeeman effects.  

All the theoretical investigations on the scattering polarization in the IR triplet of Ca\,{\sc ii} have been carried out neglecting the Zeeman splitting of the atomic levels and using either one-dimensional (1D) model atmospheres without macroscopic velocities \citep{MansoTrujillo03,MansoTrujillo10}, 1D models with radial velocities \citep{Carlin+12,Carlin+13}, or three-dimensional (3D) magneto-hydrodynamical models but neglecting the effects of symmetry breaking produced by the horizontal inhomogeneities and/or by the non-vertical velocity components of the atmospheric model used \citep{CarlinAsensio15}. Likewise, in 3D models of the solar atmosphere the polarization induced by the action of the Zeeman effect alone has been investigated using also the so-called 1.5D approximation, which neglects the impact of horizontal radiation transfer on the atomic level populations \citep{Cruz-Rodriguez+12}. Of particular interest is the conclusion of \cite{Carlin+13} that the presence of spatial gradients in the macroscopic velocity of the atmospheric plasma can have, in principle, a significant impact on the scattering polarization of the Ca {\sc ii} IR triplet.  

Here we present the main results of a full 3D radiative transfer investigation about the disk-center polarization of 
the solar Ca {\sc ii} 8542 \AA\ line, which we have carried out using the PORTA code \citep{StepanTrujillo13} and the MareNostrum supercomputer of the Barcelona Supercomputing Center (Spain). Firstly, we neglect the Zeeman effect and calculate the linear polarization due to the presence of atomic level polarization and the Hanle effect; secondly, we assume that the atomic levels are unpolarized and calculate the circular and linear polarization produced by the Zeeman effect; thirdly, we show examples of the Stokes profiles that result from the joint action of atomic level polarization and the Hanle and Zeeman effects. We consider a 3D model of an enhanced-network bipolar region  \citep[see][]{Carlsson+16}, resulting from a state-of-the-art radiation magneto-hydrodynamic simulation carried out with the Bifrost code \citep{Gudiksen+11}. In our radiative transfer calculations we take fully into account the symmetry breaking effects produced by the thermal, dynamic and magnetic structure of the model atmosphere (i.e., we do not use the 1.5D approximation). Therefore, in forward-scattering geometry (i.e., in the case of a disk-center observation) we may now find non-zero scattering polarization without the need of an inclined magnetic field. In a 1D model atmosphere with or without radial velocities, the only way to break the axial symmetry of the incident radiation field is through the presence of a magnetic field inclined with respect to the solar local vertical direction, which produces the Hanle effect. For this reason, in such a model atmosphere the Hanle effect of an inclined magnetic field creates forward scattering polarization. When a 3D model atmosphere is used, clearly the same applies to radiative transfer investigations \citep[e.g.,][]{CarlinAsensio15} that neglect the non-vertical velocity components of the model and use the 1.5D approximation. Another way of breaking the axial symmetry of the incident radiation field is through Doppler shifts in the spectral line radiation produced by macroscopic motions with {\em non-vertical} velocity components; therefore, even in an unmagnetized 1D model atmosphere we may have $Q/I$ and $U/I$ forward-scattering signals if the non-vertical velocity components are taken into account. Finally, the presence of horizontal inhomogeneities in the physical properties of the stellar atmospheric plasma (e.g., kinetic temperature and gas density) can break the axial symmetry of the incident radiation field at each point within the medium \citep[e.g.,][]{MansoTrujillo11,Stepan15}. The 3D model atmosphere we have chosen for our radiative transfer investigation has inclined magnetic fields, non-vertical velocities and horizontal inhomogeneities. The main aim of this letter is to highlight the significance of these sources of symmetry breaking in producing the linear polarization of the emergent radiation in the Ca\,{\sc ii} 8542 \AA\ line.

\section{Formulation of the 3D radiation transfer problem} 

\begin{figure}[t]
\begin{center}
\includegraphics[scale=1.0]{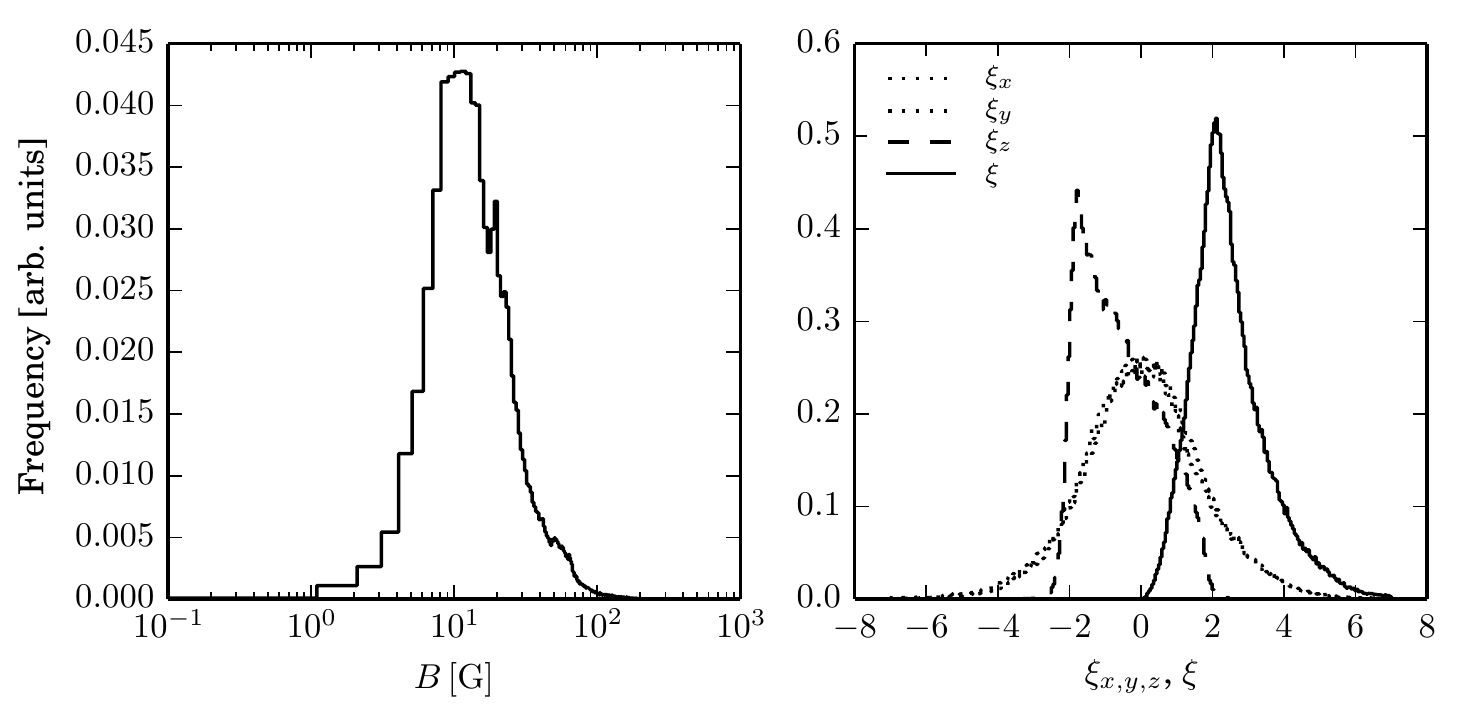}
\caption{Histograms of the magnetic field strength (left panel) and of the macroscopic velocity components in units of the line's Doppler width (right panel) obtained from the corresponding values found at the heights in the 3D model where the vertical optical depth at the center of the 8542\,\AA\ line is unity.  The solid curve in the right panel corresponds to the modulus of the velocity vector.
}
\label{figure-1}
\end{center}
\end{figure}

Fig.~\ref{figure-1} gives information on the magnetic field strength and the macroscopic velocity 
at the heights in the 3D model atmosphere that delineate the corrugated surface where the vertical optical depth at the center of the 8542\,\AA\ line is unity. Note that the magnetic strength varies between about 1 and 100 gauss and that the most likely macroscopic velocity is 2.5 km/s, which gives rise to a Doppler shift in the 8542\,\AA\ line equal to two times its thermal Doppler width.

The scattering polarization signals in the Ca {\sc ii} 8542 \AA\ line are dominated by the atomic polarization of its lower (metastable) level \citep{MansoTrujillo03,MansoTrujillo10}. As shown in figure 8 of \cite{MansoTrujillo10}, for magnetic strengths similar or larger than 10 G the scattering polarization signals of the 8542 \AA\ line are in the saturation regime of the Hanle effect (i.e., they are sensitive only to the orientation of the model's magnetic field). This implies that there should be no significant forward-scattering signals in regions of the 3D model having predominantly vertical magnetic fields stronger than 10 G, because in the Hanle saturation regime any quantum coherence between different $M$ sublevels is zero in the magnetic field reference frame and for line of sights parallel to the magnetic field vector such coherences are the only source of line scattering polarization. On the other hand, the Doppler shifts associated to the macroscopic velocities of the 3D model are a significant fraction of the line widths, so that following \cite{Carlin+13} we may expect a non-negligible impact on the scattering polarization of the emergent spectral line radiation. 

To solve the above-mentioned 3D radiation transfer problem we have applied the PORTA code \citep{StepanTrujillo13}, which 
solves jointly the radiative transfer equations for the Stokes parameters and the statistical equilibrium equations for the multipolar components of the atomic density matrix associated to each atomic level of total angular momentum $J$, assuming complete frequency redistribution and neglecting quantum interference between the magnetic sublevels pertaining to different $J$-levels \citep[see the multilevel model atom equations in Chapter~7 of][]{LL04}. We have used the 5-level model atom detailed in figure 1 and table 1 of \cite{MansoTrujillo10}, including the population transfer collisional rates for all the transitions and the alignment transfer collisional rates for the allowed transitions. \cite{MansoTrujillo10} and \cite{BelluzziTrujillo11} argue why this theoretical approach is expected to be suitable for modeling the intensity and polarization observed in the Ca\,{\sc ii} IR triplet.

It is important to note that we have carried out the radiative transfer calculations without using any \mbox{ad-hoc} line fitting parameter (i.e., we have not used microturbulent velocities). We point out that the Doppler core of the spatially-averaged intensity profile of the 8542\,\AA\ line calculated for the case of a disk center observation is approximately a factor 0.65 narrower than the observed one \citep[see the atlas of][]{Neckel99}, but both profiles have nearly similar line-center depths. This implies that the theoretical Zeeman polarization signals shown here are probably overestimated. The fact that the intensity profiles calculated in the 3D atmospheric model are narrower than the observed ones suggests that the model lacks some of the dynamics present in the real Sun. A different  dynamical activity is likely to change the scattering line polarization signals. Whether the dynamics of the real solar atmosphere has a significant influence on the scattering polarization of the Ca {\sc ii} IR triplet will depend on the particular realization of the velocity field in a fully realistic 3D model.

\section{Results}

\subsection{3D versus 1.5D}

\begin{figure}[t]
\begin{center}
\includegraphics[scale=0.65]{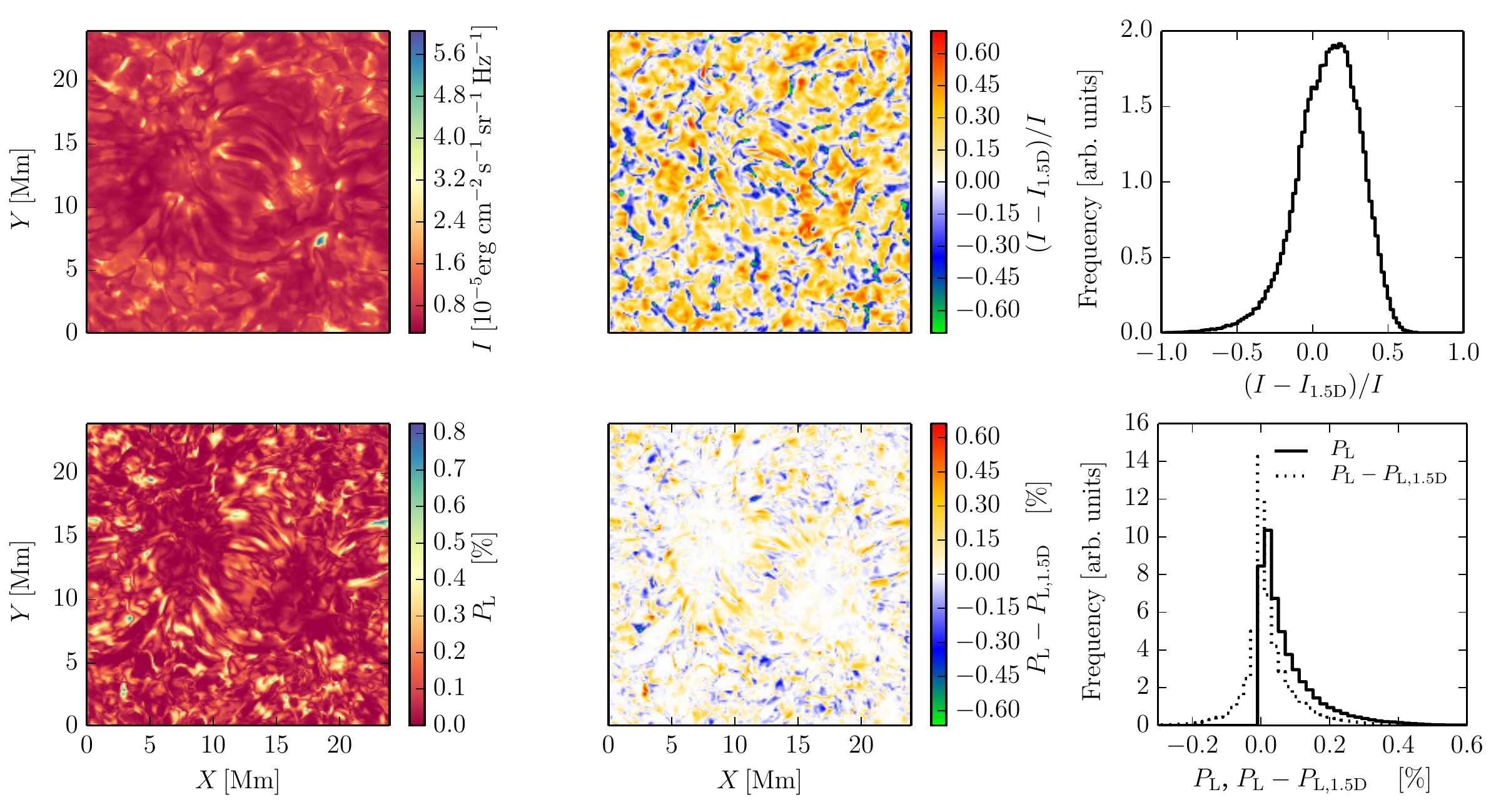}
\caption{The intensity (upper left panel) and the total fractional linear polarization (bottom left panel) 
at the center of the Ca\,{\sc ii}\,8542\,\AA\ line, calculated by solving the scattering polarization problem in the 3D atmospheric model.
The other panels quantify the errors in the results obtained with the 1.5D approximation. The upper right panel shows a histogram of the fractional errors in the intensities corresponding to each point of the field of view, while the bottom right panel shows histograms of the $P_{\rm L}$ values and of the errors in the total fractional linear polarization signals. The middle panels show the spatial distribution of such errors.
}
\label{figure-2}
\end{center}
\end{figure}

Figure \ref{figure-2} provides information about the effects of horizontal radiative transfer on the intensity (upper panels) 
and the total fractional linear polarization ($P_{\rm L}=\sqrt{Q^2+U^2}/I$; lower panels) at the center of the 8542\,\AA\ line. The central and the right panels quantify the errors made when solving the radiative transfer problem applying the so-called 1.5D approximation (i.e., assuming that each column of the 3D model can be considered as if it were an isolated 1D plane-parallel model). 

The emergent intensity of the spectral line radiation depends mainly on the variation of the total population of the upper and lower line  levels along the line of sight. The upper left panel shows the line-center intensity pattern that results from our full 3D solution.  
The right panel indicates that at the spatial resolution of the 3D model the most typical fractional error in the line-center intensity calculated with the 1.5D approximation is about 20\%, and that it can exceed more than 50\% in some points of the field of view.
The spatial distribution of such errors can be seen in the upper middle panel. Given that $V\,{\sim}\,{\rm d}I/{\rm d}{\lambda}$, the 1.5D approximation is also questionable when it comes to an accurate calculation of the circular polarization induced by the Zeeman effect.  Concerning the spatially-averaged spectral line intensities (not shown here), we point out that the differences between the full 3D and 1.5D results are not very significant.   

The scattering polarization of the emergent spectral line radiation is determined by the variation of the atomic level polarization 
along the line of sight under consideration. The bottom left panel of Fig.~\ref{figure-2} shows the total fractional linear polarization at the center of the 8542\,\AA\ line, as it results from our 3D radiative transfer calculation. The middle and the right panels show that the differences between the 3D and the 1.5D results are of the order of the true (3D) $P_{\rm  L}$-values themselves; this demonstrates that a full 3D calculation is crucial for understanding the scattering polarization signals of the Ca\,{\sc ii} IR triplet, even for the disk center observation case considered in this letter.

\subsection{Sensitivity of the line scattering polarization to the Hanle effect, to the macroscopic velocities, and to the spatial resolution of the simulated observation}

\begin{figure}[t]
\begin{center}
\includegraphics[scale=0.75]{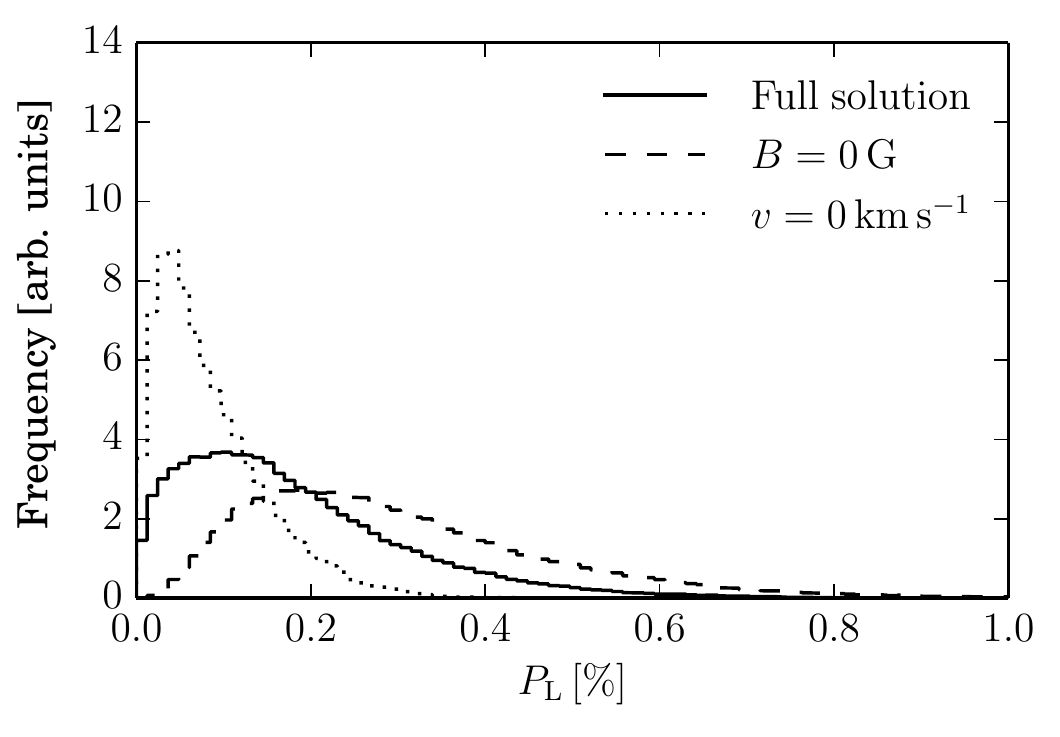}
\includegraphics[scale=0.75]{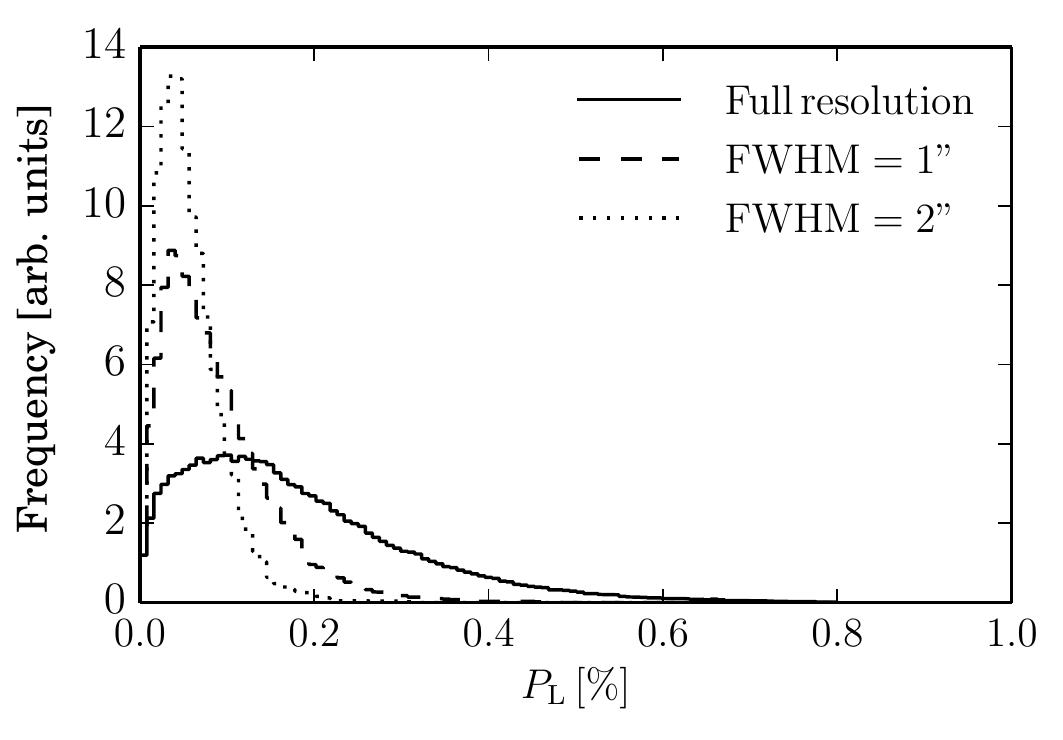}
\caption{Histograms of the distribution of the amplitudes of the total fractional linear polarization signal in the Ca\,{\sc ii} 8542\,\AA\ line 
across the field of view of the 3D model surface. As discussed in the text, the left panel shows the impact of the Hanle effect and of the 
macroscopic velocities, while the right one the reduction of the polarization amplitudes as the spatial resolution of the simulated observation is reduced (we do this by convolving the emergent Stokes profiles with gaussians having the indicated full widths at half maximum).  }
\label{figure-3}
\end{center}
\end{figure}

The linear polarization signals caused by the presence of atomic polarization in the levels of the Ca\,{\sc ii} IR triplet are, in general, sensitive to the Hanle effect and to the macroscopic velocities of the atmospheric plasma \citep{MansoTrujillo10,Carlin+12,Carlin+13}. They are also sensitive to the spatial resolution of the observation. We illustrate this by means of the full 3D results of Fig.~\ref{figure-3}, which shows  histograms of the {\em amplitudes} of the total fractional linear polarization ($P_{\rm L}$) signals in the 8542\,\AA\ line. It is important to note that Fig.~\ref{figure-3} shows the maximum amplitudes of the $P_{\rm L}$ signals, instead of the $P_{\rm L}$ values shown in the bottom left panel of Fig.~\ref{figure-2} for the laboratory line-center wavelength. 

There are several interesting results in the left panel of Fig.~\ref{figure-3}. The influence of the Hanle effect produced by the magnetic field of the 3D model can be clearly seen by comparing the dashed and solid curves. Both cases include the impact of the macroscopic velocities, but while the dashed curve corresponds to the zero magnetic field case the solid curve was obtained from the $P_{\rm L}$ signals computed taking into account the model's magnetic field. The impact of the macroscopic velocities of the 3D model can be appreciated by comparing the solid curve just mentioned with the dotted one, which corresponds to the zero velocity case. In summary, the macroscopic velocities increase the total fractional linear polarization amplitudes (dotted vs. solid curves), while the magnetic field of the 3D model produces depolarization at most of the locations (dashed vs. solid curves). 

Finally, the right panel of Fig.~\ref{figure-3} illustrates what happens when the spatial resolution of the simulated spectropolarimetric observations is reduced, in contrast with the full-resolution case characterized by a grid spacing of about 0.06".

\subsection{Hanle versus Zeeman signals}

With Figs. 4 and 5 we aim at providing information useful to understand that, in general, the linear polarization signals of the Ca\,{\sc ii} IR triplet have significant contributions coming from both the presence of atomic level polarization and the Zeeman splitting of the atomic levels. Hereafter, we will use the term ``Hanle signals" to refer to those resulting from atomic level polarization. 

\begin{figure}[t]
\begin{center}
\includegraphics[scale=0.8]{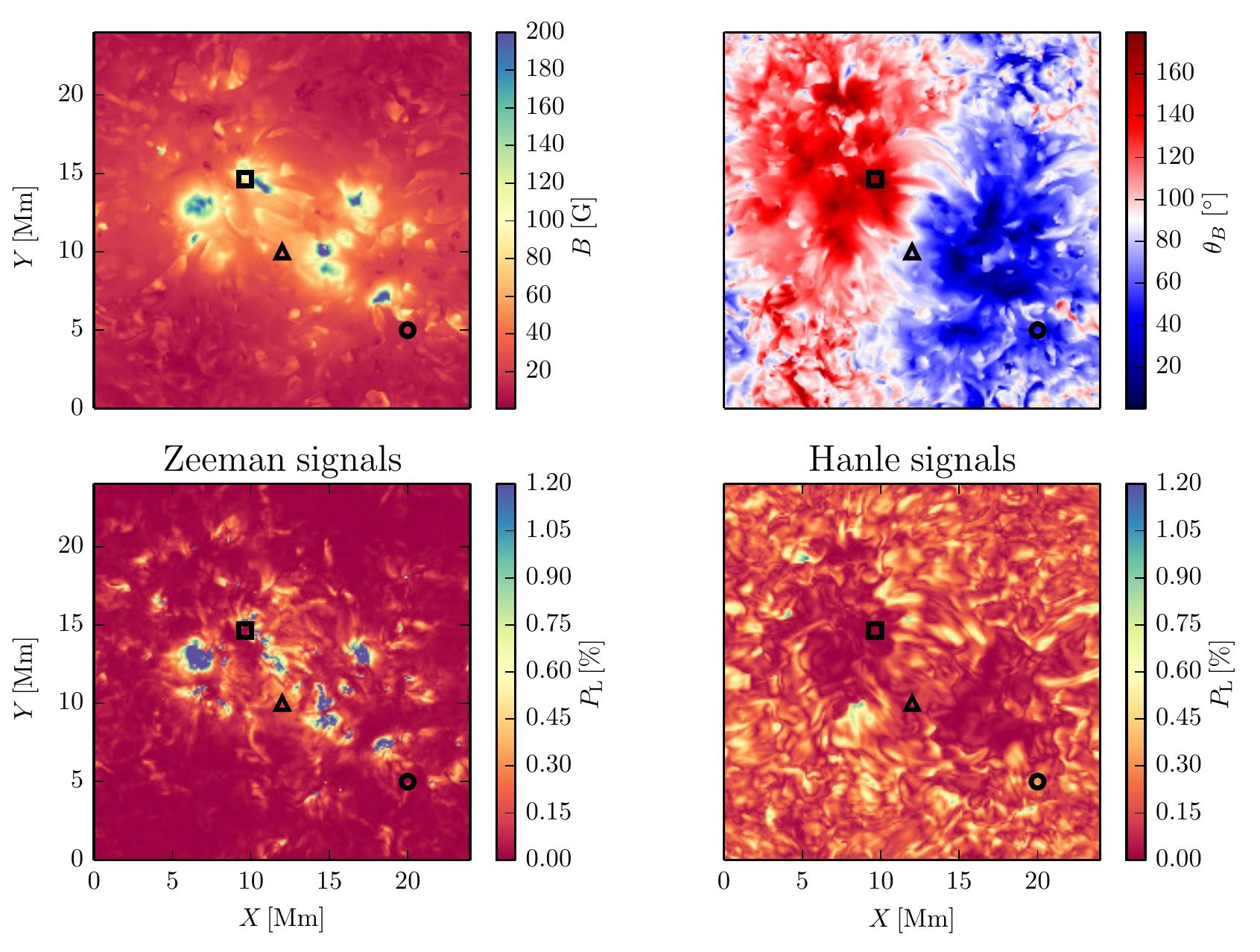}
\caption{Spatial distribution of the linear polarization amplitudes caused only by atomic level polarization (lower right panel) or by the Zeeman effect without atomic level polarization (lower left panel) in the Ca\,{\sc ii} 8542\,\AA\ line. The three symbols indicate the spatial points at which Fig.~\ref{figure-5} shows the ensuing Stokes profiles. The upper panels show the model's magnetic field strength (left) and the magnetic field inclination (right) throughout the corrugated surface where the vertical optical depth at the center of the Ca\,{\sc ii} 8542\,\AA\ line is unity.    
}
\label{figure-4}
\end{center}
\end{figure}

Fig.~\ref{figure-4} shows maps of the amplitudes of the total fractional linear polarization signals caused only by the atomic polarization of the Ca\,{\sc ii}\,8542\,\AA\ levels (right panel) and by the Zeeman effect without atomic level polarization (left panel). Qualitatively speaking, the figure indicates that in the regions of the field of view where the Zeeman signals are the largest (i.e., at the two opposite-polarity localized magnetic patches where the magnetic field is stronger than 100 G and slightly inclined) the Hanle signals are relatively weak, while in the regions where the Hanle signals reach their maximum amplitudes (i.e., outside such magnetic patches) the Zeeman signals are much weaker or negligible.  

\begin{figure}[t]
\begin{center}
\includegraphics[scale=0.3]{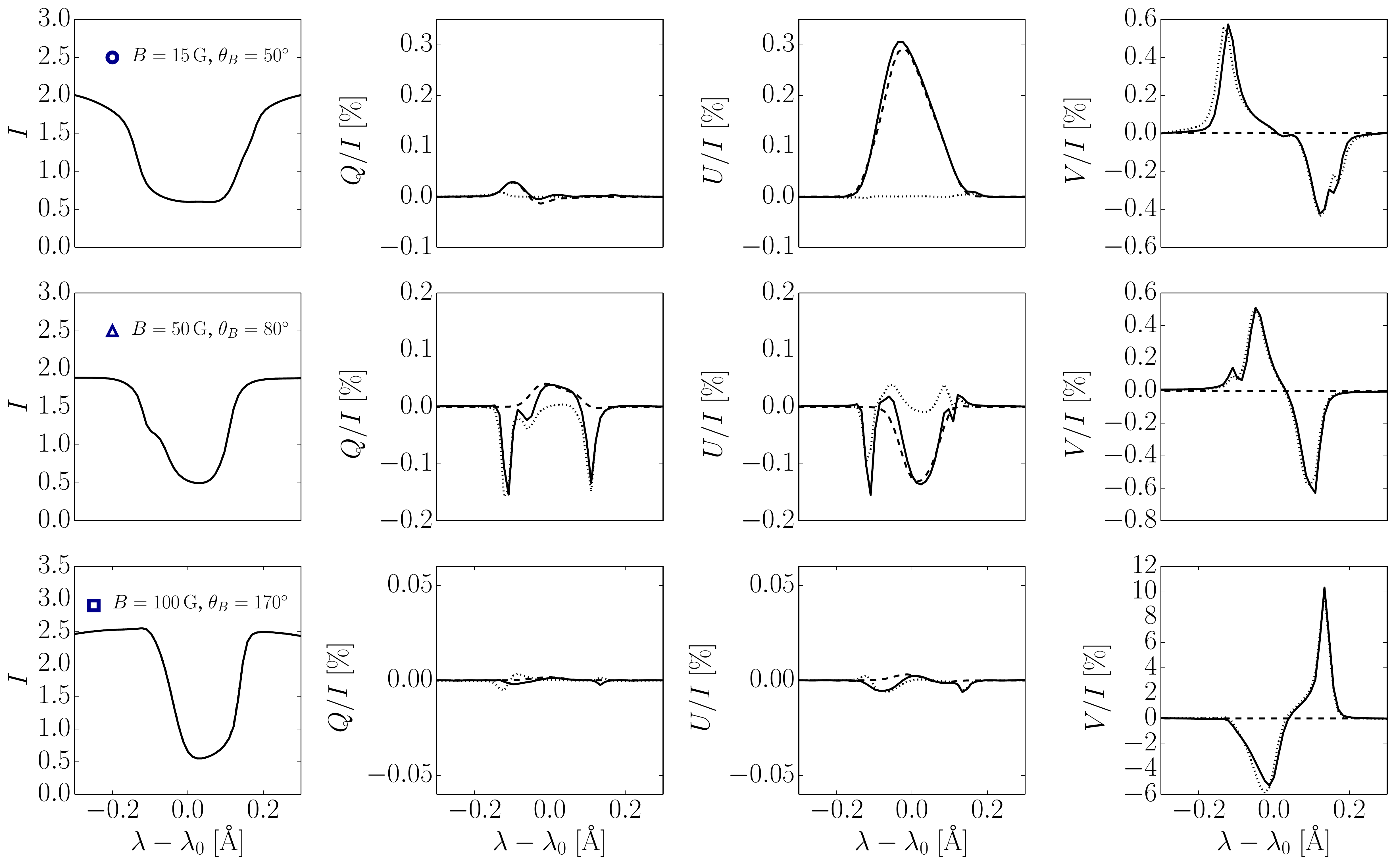}
\caption{The Stokes profiles of the Ca\,{\sc ii} 8542 \AA\ line corresponding to the spatial points indicated in Fig.~\ref{figure-4}, obtained  taking into account only the contribution of atomic level polarization and the Hanle effect (dashed  curves), only the Zeeman effect without atomic level polarization (dotted curves), and the joint action of atomic level polarization and the Hanle and Zeeman effects (solid curves). Stokes $I$ is given in units of $10^{-5}\,{\rm ergs\,cm^{-2}\,s^{-1}\,sr^{-1}\,Hz^{-1}}$.
}
\label{figure-5}
\end{center}
\end{figure}

The upper panels of Fig.~\ref{figure-5} show an example of emergent Stokes profiles corresponding to a spatial location where the linear polarization of the 8542\,\AA\ line is dominated by the contribution of atomic level polarization. Likewise, the bottom panels of Fig.~\ref{figure-5} show the Stokes profiles of a spatial location where the linear polarization is dominated by the Zeeman effect. In other spatial locations, the Hanle and Zeeman signals are of the same order of magnitude, as illustrated in the middle panel of Fig.~\ref{figure-5}. As expected, at such spatial locations the $Q/I$ and $U/I$ line center signals are dominated by the atomic level polarization and the Hanle effect, while the wing signals are produced by the Zeeman effect. 
In any case, we point out that at most of the points of the field of view the Hanle signals dominate (see the bottom panels of Fig~\ref{figure-4}). Finally, note that the right panels of Fig.~\ref{figure-5} show the circular polarization caused by the Zeeman effect, without (dotted curves) and with (solid curves) atomic level polarization, and that for the spatially resolved situation considered in this section the ensuing Stokes-$V$ amplitudes are significant at practically all spatial locations.

\section{Concluding comments}

Understanding the polarization of the IR triplet of Ca\,{\sc ii} is important, both to interpret correctly 
the spectropolarimetric observations that can be carried out with the available telescopes and for designing the most suitable instruments for future investigations with ground-based and spaceborne facilities. With this scientific motivation, we have solved the problem of the generation and transfer of polarization in the Ca\,{\sc ii} IR triplet using a 3D model of the solar atmosphere, taking fully into account the effects of horizontal radiative transfer.

In the 3D model we have considered, the Zeeman effect tends to dominate the linear polarization of the 8542 \AA\ line  
in the localized patches of opposite magnetic polarity where the magnetic field is relatively strong ($B>100$ G)   
and slightly inclined, while in weakly magnetized regions ($B<100$ G)  
the linear polarization is mainly determined by the presence of atomic level polarization and the Hanle effect. It is however important to note that at many points of the model's surface both sources of spectral line polarization contribute to the linear polarization of the Ca\,{\sc ii}~8542 \AA\ line, with the line center signals determined by the atomic level polarization and the line wing signals by the Zeeman effect. The circular polarization is always dominated by the Zeeman effect.

In general, the atomic level polarization makes a significant contribution to the linear polarization of the Ca\,{\sc ii}~8542 \AA\ line. Since the polarization of the Ca\,{\sc ii} levels is caused by radiative transitions, it is sensitive to the thermal and dynamic structure of the solar atmosphere, in addition to its magnetic structure through the Hanle effect. In the 3D model atmosphere we have considered, the line Doppler shifts resulting from the macroscopic velocities significantly modify the fractional total linear polarization amplitudes.

As shown in Fig. \ref{figure-3}, the magnetic field of the 3D model produces mainly depolarization. The fact that we have Hanle depolarization even in the disk-center case is a logical consequence of the result illustrated in Fig.~\ref{figure-2}, namely the unsuitability of the 1.5D approximation given the significant breaking of the axial symmetry of the pumping radiation field caused by the horizontal inhomogeneities in the model's chromosphere.               
    
Our results lead us to emphasize that, in general, a reliable radiative transfer modeling of the polarization in the Ca\,{\sc ii} IR triplet requires taking into account the joint action of anisotropic radiation pumping (which produces atomic level polarization, even in the presence of relatively strong fields) and the Hanle and Zeeman effects in realistic 3D models of the solar chromosphere.  

\acknowledgements

We are grateful to Tanaus\'u del Pino Alem\'an (HAO) and Edgar Carl\'\i n (IRSOL) for carefully revising the manuscript and for scientific discussions. 
The calculations were carried out with the MareNostrum supercomputer of the Barcelona Supercomputing Center (National Supercomputing Center, Barcelona, Spain), and we gratefully acknowledge the computing grants. 
We are also grateful to the European Union COST action MP1104 for financing scientific missions at the IAC that facilitated the development of this investigation. 
Financial support by the Grant Agency of the Czech Republic through grant \mbox{16-16861S} and project \mbox{RVO:67985815}, as well as by the Spanish Ministry of Economy and Competitiveness through projects \mbox{AYA2014-55078-P} and \mbox{AYA2014-60476-P} is gratefully acknowledged.

\end{document}